\input harvmac.tex
\Title{\vbox{\baselineskip14pt\hbox{HUTP-97/A034}\hbox{hep-th/9707131}}}
{Geometric Origin of Montonen-Olive Duality}
\bigskip\vskip2ex
\centerline{Cumrun Vafa}
\vskip2ex
\centerline{\it  Lyman Laboratory of Physics, Harvard
University}
\centerline{\it Cambridge, MA 02138, USA}
\vskip .3in

We show how $N=4, D=4$ duality of Montonen and Olive can be derived
for all gauge groups
using geometric engineering in the context of type II strings, where
it reduces to T-duality.
The derivation for the non-simply laced cases involves the use of
some well known facts about orbifold conformal theories.

\Date{\it {July 1997}}

\newsec{Introduction}
The aim of this note is to show how the celebrated Montonen-Olive duality
\ref\mol{C. Montonen and D. Olive, Phys. Lett. {\bf B72} (1977) 117
\semi P. Goddard, J. Nyuts and D. Oive, Nucl. Phys. {\bf B125} (1977) 1 \semi
H. Osborn, Phys. Lett. {\bf B83} (1979) 321.}\
for all $N=4$ gauge theories in $D=4$
can be
derived by geometric engineering in the context of type II strings,
where it reduces to T-duality.
Even though by now there is a lot of evidence for the Montonen-Olive
duality (see e.g. \ref\revom{A. Sen, Phys. Lett. {\bf B329} (1994) 217\semi
C. Vafa and E. Witten, Nucl. Phys. {\bf B431} (1994) 3\semi
M. Bershadsky, A. Johansen, V. Sadov and C. Vafa, Nucl. Phys.
{\bf B448} (1995) 166\semi
J. Harvey, G. Moore and A. Strominger, Phys.Rev. {\bf D52} (1995) 7161\semi
L. Girardello, A. Giveon, M. Porrati, A. Zaffaroni,
Nucl. Phys. Proc. Suppl. {\bf 46}(1996) 75.})\ there is no
derivation of this duality.  Even with the recent advances
in our understanding of dynamics of string theory the derivation
of this duality is not yet complete.  The aim of this note is to
fill this gap.
The approach we will follow
is in the context of type II compactifications
and is quite general and provides a unified
approach to all gauge groups.
   Moreover
we gain an understanding of how the field theory duality
 works by relating it to well understood
perturbative symmetries (T-dualities) of strings.\foot{For
the case of $SU(n)$ gauge group there is another
approach suggested in \ref\strom3{A. Strominger,
Phys. Lett. {\bf B383} (1996) 44
\semi M. Douglas, hepth/9604198.}\
which uses the Hull-Townsend $SL(2,{\bf Z})$ duality of type IIB
strings \ref\hult{C. Hull and P. Townsend, Nucl. Phys. {\bf B438} (1995) 109.}.
 In this approach one considers the theory of $n$ parallel $D3$
branes, and uses the fact that the $SL(2,{\bf Z})$ symmetry maps $D3$ branes
 back to itself.  It would be interesting to see if this approach
can be generalized to all gauge groups.}

\newsec{Montonen-Olive Duality}
Let us recall what the Montonen-Olive duality is:
We consider $N=4$ supersymmetric gauge theories in $D=4$. Consider
gauge group $G$, with gauge coupling constant $g$.  Then
the Montonen-Olive duality suggests that this theory is equivalent
to $N=4$ gauge theory with a dual gauge group ${\widehat G}$ and coupling
constant $g'\propto 1/g$ where the electric and magnetic degrees
of freedom are exchanged.
  For the $A_n,D_n,E_{6,7,8},F_4,G_2$ gauge groups
${\widehat G}$
is again the same gauge group\foot{Here we do not
pay attention to global issues and limit ourselves
to Lie algebras.  However,  more precisely
the Montonen-Olive duality suggests that the weight lattice
of the dual group is dual to the weight lattice of the original
group. For example if we start with $SU(n)$ the dual group is $SU(n)/Z_n$.}.
On the other
hand the Montonen-Olive duality exchanges the $B_n$ and $C_n$
gauge groups.

\newsec{Basic idea of geometric engineering of $N=4$ theories}
We now review the basic idea of geometric engineering of $N=4$
theories in four dimensions in the context of type IIA strings.
We will start with the simpler case of the simply laced groups.

\subsec{Simply laced cases: A-D-E}

Consider type IIA strings propagating on an ALE space of A-D-E type.
Then as is well known this gives rise in six dimensions to an $N=2$
theory with A-D-E gauge group, where the charged gauge particles are obtained
by wrapping D2 branes around 2-cycles of the ALE space, and the uncharged
(Cartan) gauge fields arise from decomposition of the 3-form gauge potential
in terms of the harmonic 2-forms corresponding to the compact
cycles of the ALE space.  If we compactify on a $T^2$ from 6 down to 4
dimensions, the gauge coupling in $d=4$ is proportional to the volume of
$T^2$.  The inversion of the coupling constant in the four dimensional
theory amounts to the T-duality volume inversion symmetry of type IIA strings
on a $T^2$.  This establishes the Montonen-Olive self-duality
of A-D-E gauge groups by reducing it to string T-duality\foot{The
connection between S-duality and T-duality in string theories
was first suggested
in \ref\duff{M. Duff, Nucl. Phys. {\bf B442} (1995) 47.}. In the compact
version of the above construction,
it was noted in \ref\wit{E. Witten, Nucl. Phys. {\bf B443} (1995) 85.
}\ to be a consequence of string-string duality.
Note that here we are taking a slightly different view by not utilizing
string-string duality, and just using facts about D-branes which
were established later, thanks to the observations in \ref\strom{
A. Strominger, Nucl. Phys. {\bf B451} (1995) 96.}
and \ref\pol{J. Polchinski, Phys. Rev. Lett. {\bf 75} (1995) 4724.}.  See in
particular \ref\bsv{M. Bershadsky, V.
Sadov and C. Vafa, Nucl. Phys. {\bf B463} (1996) 398
\semi Nucl. Phys. {\bf B463} (1996) 420.}\ref\dougl{M. Douglas,
hepth/9612126.}.}.

\subsec{The non-simply laced cases}
To geometrically engineer non-simply laced
 gauge groups in four dimensions we follow
the idea in \ref\aspg{P. Aspinwall and M. Gross, Phys.Lett. {\bf B387} (1996)
735.} (see also \ref\sixau{Bershadsky et. al., Nucl. Phys. {\bf B481}
(1996) 215.}),
by using outer automorphisms of the simply laced groups.
We consider an A-D-E ALE space, and compactify on an extra
circle and identify translation along $1/n$-th of the circle
by a specific outer automorphism ${\bf Z_n}$ symmetry acting on an ALE space.
In other words we consider the 5-dimensional space
$$M={{\rm ALE}\times S^1\over Z_n}$$
as the background, where ${\bf Z_n}$ acts simultaneously as an outer
automorphism
of the $ALE$ space and an order $n$ translation on the circle.
The relevant symmetries for the various non-simply laced groups are:
$${(D_{n+1},{\bf Z_2})\rightarrow B_n}$$
$$(A_{2n-1},{\bf Z_2})\rightarrow C_n$$
$$(E_6,{\bf Z_2})\rightarrow F_4$$
\eqn\outer{(D_4,{\bf Z_3})\rightarrow G_2}
where in the $D$ case the $Z_2$ exchanges the two
end nodes, for $A_{2n-1}$ and $E_6$ case it flips the Dynkin diagram through
the middle node and for $D_4$ it permutes the three outer nodes.

The strategy we will follow is to show that
type IIA on
$$M={{\rm ALE}\times S^1(R)\over {\bf Z_n}}$$
is equivalent to type IIB strings on
$$\widehat M={\widehat{{\rm ALE}}\times S^1(n/R)\over {\bf Z_n}}$$
where the ${\bf Z_n}$'s are according to \outer\ and the
dual $\widehat{{\rm ALE}}$ corresponds to the same ALE for
$E_6$ and $D_4$ but exchanges $A_{2n-1}$ and $D_{n+1}$ ALE spaces.
Compactifying further on another circle and using the $R\rightarrow
1/R$ symmetry on the other circle converts the theory back from type IIB
to type IIA, and we will thus have established Montonen-Olive
duality using T-duality, by showing the type IIA string equivalence of
\eqn\eqta{{{\rm ALE}\times S^1(R)
\over {\bf Z_n}}\times S^1(R')={\widehat{{\rm ALE}}\times S^1(n/R)\over {\bf
Z_n}}\times
S^1(1/R')} %
\newsec{Some facts about orbifold CFT's}
In this section we will review some facts about ${\bf Z_n}$ orbifold
conformal field theories that we will use in the next section.

Let $C$ denote a conformal theory with a ${\bf Z_n}$ discrete symmetry.
We can consider orbifolding this theory with the ${\bf Z_n}$ symmetry.
Let us recall some aspects of how this works \ref\dhvw{Dixon et. al., Nucl.
Phys. {\bf B274} (1986) 285.}
(see also the review article \ref\gins{P. Ginsparg,``Applied Conformal
Field Theory,'' proceedings of Les Houches Lectures 1988,
Elsevier Science Publishers B.V. (1989). }).
There are $n$ twisted sectors, labeled by an integer $r$ mod $n$,
which are the sector of strings closed up to the ${\bf Z_n}$
action.  Let us denote the Hilbert of each sector by $C_r$.
Moreover, we can decompose each sector according to how the ${\bf Z_n}$
acts on that sector.  Let $C_r^s$ denote the subsector of the $r$-th
twisted sector which transforms according to ${\rm exp}[2\pi is/n]$
where $s$ is also defined mod $n$.  We can associate another
$\bf {{\tilde Z}_n}$ symmetry by using the grading of the twist sector,
i.e. by considering $C_r^s$ to transform as ${\rm exp}[2 \pi i r/n]$.
As is well known, the Hilbert space of conformal theory $C/{\bf Z_n}$ is
obtained
by considering the $Z_n$ invariant pieces of each sector, i.e.
$${\tilde C}={C\over {\bf Z_n}}=\sum_r C_r^0$$
It is easy to see (see \gins\ for  a review) that we can
mod ${\tilde C}$ by ${{\bf{\tilde Z}_n}}$ and recover $C$ back, i.e.
$${{\tilde C}\over {\bf{\tilde Z}_n}}=C.$$
In fact the $s$-th twisted sector of the ${\tilde C}/{\bf {\tilde Z}_n}$
can be identified with $C_r^s$, and projecting to the
${\bf Z_n}$ invariant sector means keeping $\sum C_0^s$ which
is the definition of the $C$ theory Hilbert space.
  Thus the two theories $C$ and
$\tilde C$ are on the same footing:
out of the $n^2$
subsectors $C_r^s$, exchanging $C\leftrightarrow {\tilde C}$ amounts
to exchanging $r\leftrightarrow s$.

Now suppose we have two conformal theories $C_1$ and $C_2$
each with a ${\bf Z_n}$ symmetry.  Consider orbifolding
with a single ${\bf Z_n}$ which acts on both at the same
time, on one as the generator of the original
${\bf Z_n}$ and on the other, as the inverse generator.  Then it is easy
to see that
\eqn\imp{{C_1\times C_2\over \bf Z_n}={\tilde C_1\times \tilde C_2\over
{\bf {\tilde Z}_n}}}
In fact the Hilbert space for both theories is given by
$$\sum_{r,s} C_{1,r}^s {C_{2,-r}^{-s}}$$
Note the symmetrical role $r$ and $s$ play in the above
expression, which is a reflection of the equivalence \imp,
as $C\leftrightarrow {\tilde C}$ amounts to $r\leftrightarrow s$.
The equivalence \imp\ is what we will use to prove the duality we
are after.  In fact this is exactly of the form of the equivalence
\eqta\ that we wish to prove (the second circle plays no major role), where
$C_1, \tilde C_1$
are to be identified with the ALE theory and its dual and
$C_2, \tilde C_2$ with the $S^1(R)$ theory and its dual
$S^1(n/R)$.  The fact that $S^1(R)/{\bf Z_n}=S^1(n/R)$ is straight
forward.  In fact, by definition of the ${\bf Z_n}$ action
on the circle we have
$$S^1(R)/{\bf Z_n}=S^1(R/n).$$
Applying the standard T-duality on this we get $S^1(n/R)$ which
is thus the dual theory ${\tilde C_2}$.  Note that the ${\bf
{\tilde Z}_n}$ acting on ${\tilde C_2}$ is a translation of order $n$
on this dual circle.  In fact $S^1(n/R)/{\bf {\tilde Z}_n}=S^1(1/R)$
which by the standard T-duality is equivalent to $S^1(R)$.
So all we are left to do to complete the proof of Montonen-Olive
duality is to prove $C_1,{\tilde C_1}$ are dual
conformal theories, as predicted by the duality.  This we will do in the next
section.

\newsec{Proving the Duality}
We complete the proof of duality in this section by showing the following
dualities of CFT on ALE spaces:
$$A_{2n-1}/{\bf Z_2}=D_{n+1}\qquad D_{n+1}/{\bf Z_2}=A_{2n-1} \qquad
(C_n\leftrightarrow B_n)$$
$$E_6/{\bf Z_2}=E_6 \qquad (F_4\leftrightarrow F_4)$$
\eqn\givn{D_4/{\bf Z_3}=D_4\qquad (G_2\leftrightarrow G_2 )}
  This would identify the $C_1,\tilde C_1$ in
\imp\ with the appropriate dual needed in
\outer , which completes the proof of Montonen-Olive duality
in accordance
with \eqta .

In order to do this we need to recall the $N=2$ superconformal
theories associated with string propagation on
ALE spaces.  It was shown in \ref\ov{H. Ooguri and
C. Vafa,
Nucl.Phys. {\bf B463}(1996) 55.}\
that this is described by the $N=2$ Landau-Ginzburg theory with
superpotential
$$W=W_{ADE}(x,y,z)-  t^{-d}$$
where $W_{ADE}$ denotes the $ADE$ singularity and $d$ is the
dual coxeter number of the corresponding singularity:
$$W_{A_{n-1}}=x^n+y^2+z^2\qquad d=n$$
$$W_{D_n}=x^{n-1}+xy^2+z^2\qquad d=2n-2$$
$$W_{E_6}=x^3+y^4+z^2\qquad d=12$$
$$W_{E_7}=x^3+xy^3+z^2\qquad d=18$$
$$W_{E_8}=x^3+y^5+z^2\qquad d=30$$
 (we also
need to mod out by ${\rm exp}(2\pi i J_0)$ which is the generator
of a ${\bf Z_d}$--this however will not play a major
role in the following).
As a superconformal theory, the part corresponding to $W_{ADE}$
is equivalent to the corresponding $N=2$ minimal models, as was shown in
\ref\vw{C. Vafa and N. Warner, Phys. Lett. {\bf B218} (1989) 51\semi
E. Martinec, Phys. Lett. {\bf B217} (1989) 431.}, and the part corresponding
to $t^{-d}$ term is a Kazama-Suzuki model based on $SL(2)$ group
(at level $d+2$).  The symmetries we are interested in modding
out act only on the $x,y,z$ variables, thus the latter superconformal
theory plays no role.  So all we need to show is that the orbifold
of minimal $N=2$ superconformal theories for $A_{2n-1}$,$D_{n-1}$,
$E_6$ and $D_4$ behave as expected from \givn .  This fact is actually
well known, and can be readily derived since the minimal conformal
theories are very well known.  Here
we shall review it for completeness and present its derivation along the
lines suggested in \ref\schm{R. Schimmrigk and M. Lynker, Phys. Lett.
{\bf B249} (1990) 237.}.

Consider $A_{2n-1}$ minimal model:
$$W_{A_{2n-1}}=x^{2n}+y^2+z^2.$$
The relevant outer automorphism ${\bf Z}_2$ acts as
$$x\rightarrow -x,\quad y\rightarrow -y, \quad z\rightarrow z$$
Thus we introduce the invariant variables
$$\tilde x = x^2\quad
\tilde y=y/x\quad \tilde z=z$$
 (which have been chosen
to keep the Jacobian of transformation constant).  Then we obtain\foot{
The result of this ${\bf Z_2}$ modding out for conformal
theories associated to ALE is {\it not} the one
that would be naively expected based on the geometry of $ALE$ space
for particle theories, where one would end up instead with $D_{n+2}$.}
$$W_{A_{2n-1}}/{\bf Z_2}={\tilde x}^n+\tilde x\tilde y^2+\tilde z^2
=W_{D_{n+1}}$$
To go the reverse, it is of course true on general
grounds discussed above that there is a ${\bf Z_2}$
acting on $D_{n+1}$ which gives back $A_{2n-1}$.  However it is not
apriori obvious why it should be the one corresponding to the outer
automorphism of $D_{n+1}$. To accomplish this we show directly that
the outer automorphism ${\bf Z_2}$ leads back to the $A_{2n-1}$ theory.
We have
$$W_{D_{n+1}}=x^n+xy^2+z^2$$
and the outer automorphism ${\bf Z_2}$ acts by
$$x\rightarrow x,\quad y\rightarrow -y,\quad z\rightarrow -z.$$
We introduce the new invariant variables
$$\tilde x=x\quad \tilde y =y^2\quad \tilde z=z/y$$
which leads to
$$W_{D_{n+1}}/{\bf Z}_2=\tilde x^n+\tilde y(\tilde x+\tilde z^2)=W_{A_{2n-1}}$$
(the last equality follows by shift of variables, or simply by noting
that the variation with respect to $\tilde y$ sets $\tilde x=-\tilde z^2$
which leads to $\tilde z^{2n}$).

For $E_6$ we have
$$W_{E_6}=x^3+y^4+z^2$$
and the ${\bf Z}_2$ outer automorphism is given by
$$x\rightarrow x, \quad y\rightarrow -y,\quad z\rightarrow -z.$$
We introduce the new variables
$$\tilde x=x\quad \tilde y=y^2 \quad\tilde z=z/y$$
which leads to
$$W_{E_6}/{\bf Z_2}=\tilde x^3+\tilde y^2 +\tilde y\tilde z^2=
\tilde x^3+(\tilde y +{1\over 2}\tilde z^2)^2-{1\over 4}\tilde z^4=
W_{E_6}$$
(the last equality follows by shifting $\tilde y$ by ${-1\over 2}\tilde z^2$).

For the $D_4$ case we have
$$W_{D_4}=x^3+y^3+z^2$$
and the ${\bf Z_3}$ outer automorphism is given by
$$x\rightarrow \omega x,\quad y\rightarrow \omega^{-1} y,\quad
z\rightarrow z; \qquad \omega^3=1$$
and we introduce the invariant variables (again with Jacobian of
the transformation being constant)
$$\tilde x=x^2/y\quad \tilde y=y^2/x \quad \tilde z=z$$
which leads to
$$W_{D_4}/{\bf Z_3}={\tilde x}^2\tilde y+{\tilde y}^2{\tilde x}+{\tilde z}^2=
W_{D_4}$$
(where the last equality follows by shifting ${\tilde x}$ and ${\tilde y}$).
This completes the proof.

\vglue 1cm

We would like to thank S. Katz and P. Mayr for valuable discussions.

This research is supported in part by NSF grant PHY-92-18167.
\listrefs
\end